\documentstyle[11pt,aaspp4]{article}

\slugcomment{Accepted for publication in PASP}

\lefthead{Nakajima}
\righthead{Ground-Based Interferometer}

\begin{document}

\title{Sensitivity of a Ground-Based Infrared Interferometer 
for Aperture Synthesis Imaging}

\author{Tadashi Nakajima}
\affil{National Astronomical Observatory, 2-21-1 Osawa, Mitaka,
181-8588, Japan; tadashi.nakajima@nao.ac.jp}


\begin{abstract}
Sensitivity limits of ground-based infrared interferometers 
using aperture synthesis are presented.
The motivation of this analysis is to compare
an interferometer composed of multiple large telescopes and a
single giant telescope with adaptive optics.  
In deriving these limits, perfect wavefront correction 
by adaptive optics and perfect cophasing
by fringe tracking are assumed.
We consider the case in which $n$ beams are pairwise combined
at $n(n-1)/2$ detectors ($^nC_2$ interferometer)
and the case in which all the $n$ beams
are combined at a single detector ($^nC_n$ interferometer).
Our analysis covers broadband observations by considering
spectral dispersion of interference fringes.
In the read noise limit, the $^nC_n$ interferometer with 
one dimensional baseline configuration is superior to the $^nC_2$
interferometer, while in the background limit, the advantage of
the one-dimensional $^nC_n$ interferometer is small. 
As a case study,
we compare the 
point-source sensitivities of interferometers composed of nine
10-m diameter  telescopes and a 30-m diameter single telescope
with adaptive optics between 1 and 10 $\mu$m
for 10-$\sigma$ detection in one hour.
At J and H, the sensitivities of the interferometers are
limited to 25 to 24 mag  by read noise and OH airglow background,
while that of the single telescope is limited to 28 to 26 mag by
OH airglow background. Longward of 2 $\mu$m, the sensitivities
of the interferometers and the single telescope are all limited
by instrumental thermal background. At K, the sensitivities
of the interferometers are around 23 mag, while that of the single
telescope is 27 mag. At N, the sensitivities of the interferometers
are around 10.7 mag, while that of the single telescope is 14.4 mag.

\end{abstract}




%

\section{Introduction}

Fundamental limitations of optical and infrared aperture synthesis
imaging had been an academic subject in which astronomers were 
not keenly interested. 
The analyses of sensitivities are 
 mathematically rather involved and they appeared in journals
of optics (\cite{pra89} (PK89), \cite{nak01} (NM01)) 
instead of those in astronomy.
The analyses were academic partly because they dealt with
ideal conditions in which atmospheric disturbances did not exist.
Therefore their results were applicable only to space interferometers
which were instruments in the distant future.
They were also academic partly because optical and infrared
interferometers were not considered as routinely used astronomical
instruments unlike radio interferometers in radio astronomy.

However, as pointed out by \cite{rod99} (hereafter RR99), 
the situation has changed
due to the completion of large ground-based telescopes
with 8-10 m diameters and to the planning of next-generation 
giant telescopes with
30-100 m diameters.
In the plans of the giant telescopes, they are all equipped with
adaptive optics which deliver diffraction limited images. 
The giant telescopes are always useful as optical buckets.
However for the purpose of high resolution
imaging, it is not clear whether the synthesized images from an 
interferometric array
of large 
telescopes or diffraction limited images of giant telescopes
are superior. In terms of resolution, the interferometric array
is obviously advantageous, but in terms of sensitivity or
dynamic range the quantitative comparison is more complex.
Optical/infrared interferometric
arrays with small telescopes currently in operation
or being built are dedicated to stellar astronomy.
With the advent of
an interferometric array composed of large telescopes, one essential
question is whether interferometry finally has an impact on extragalactic
astronomy. This question is directly related to the limiting magnitude.

The development of the fringe-tracking technique has enabled
to cophase a number of telescopes and a ground-based interferometric array
can be regarded as an ideal interferometer without atmospheric disturbances.
Therefore the sensitivity analyses made for space interferometers in
the past can now be applied to ground-based interferometers as well.  
Under these circumstances, it is worthwhile to highlight
the results of previous analyses of optical and infrared synthesis
imaging in an astronomical journal and apply them to 
ground-based interferometry composed of
large telescopes.

RR99 pioneered the analysis of ground-based interferometry
from the point of view of comparison with future giant telescopes.
However their analysis is not of aperture synthesis and
as we will show later that their results are not necessarily
in agreement with ours. 
Although this paper mostly utilizes previously obtained 
formulae of signal-to-noise ratios (SNRs) in PK89 and NM01,
we also present new results to cover
the full parameter space.

The beam combination geometry is a major issue in studying
the sensitivity of an interferometer.
One extreme is the $^nC_2$ interferometer, in which $n$ beams
are divided into $n(n-1)$ subbeams and they are pairwise
combined at $^nC_2 = n(n-1)/2$ detectors. There is
one detector for each baseline. The other extreme is the $^nC_n$
interferometer in which all the $n$ beams are combined
at $^nC_n = 1$ detector. 
In this paper, we present the sensitivities
of both interferometers. We consider idealized interferometers, but
the beam-combination scenario for actual interferometers may be more
complicated (\cite{moz00}).
The $^nC_2$ and $^nC_n$ schemes are called Pairwise and All-On-One
combinations in \cite{moz00}.

Baseline redundancy is a factor that affects the SNR of
a synthesis image for a given number of apertures.
Since the number of large telescopes combined to
form an interferometric array is likely to be limited, it is important to gain
$uv$ coverage by using nonredundant baselines. Throughout this 
paper, we only consider interferometers with nonredundant
baselines.

There are two types of noise which appear in aperture
synthesis, one of which is removable while the other is
intrinsic and not removable.
Sampling noise is potentially
introduced by the incompleteness of $uv$ coverage.
However, deconvolution techniques
such as CLEAN and Maximum Entropy Method (MEM) (\cite{per89}),
when applied to radio interferometric data,
appear to compensate for sampling noise.
We thus regard sampling noise as removable noise.
On the other hand, noise resulting from the photodetection process
is not removable.
The SNR of a dirty image to which deconvolution has not been
applied is limited by photodetection noise and detector read noise.
Here we regard the SNR of a dirty image 
as the SNR of the synthesized image. 

There are two different methods by which a synthesized
image can be constructed from the visibility data: inversion
without total counts and true inversion.
In the former, the zero frequency phasor is neglected, 
implying that the total number of photons in the
synthesized image is zero. 
Despite this seemingly unattractive feature, this is the
standard method in radio astronomy.
In the latter method, the van Cittert-Zernike
theorem is strictly applied and the zero frequency
phasor as well as the $n(n-1)/2$ complex
phasors is used. The image produced by this technique has the
desirable property of nonnegativity.
In the presence of additive background which overwhelms
the signal, the magnitude of the zero frequency phasor is much greater than
those of other phasors.  
The zero frequency phasor determines the DC offset level due to
thermal background in the synthesized image. Not only is this
background of little astronomical interest, it is also a major source
of Poisson fluctuations. 
In this paper, we present the results
obtained for both inversion with and without total counts.
However, in practice faint source detection is limited by OH airglow
background at J and H
and thermal background in the thermal
infrared. So we obtain those detection limits for 
inversion without total count.

The paper is organized as follows.
First we introduce two types of interferometers,
the $^nC_2$ interferometer and $^nC_n$ interferometer
with expressions for SNRs of 
the images obtained by these interferometers
in \S2 and \S3. 
We discuss limiting cases for spectrally broad light  in \S4.
In \S5, we present the sensitivities of interferometers
composed of nine 10-m telescopes in comparison with
a single 30-m telescope with adaptive optics.
We discuss further on the sensitivity analysis in \S6.

\section{$^nC_2$ interferometer}

\placefigure{fig1}

Let there be $n$ identical principal apertures from
which we derive $n$ main beams. Each beam is divided into
$n-1$ subbeams by the use of beam splitters. The resulting
$n(n-1)$ subbeams are combined pairwise at $n_b = n(n-1)/2$
detectors where $n_b$ is also the number of baselines.
The fringe pattern is formed at the focal plane of
each detector which is either a one dimensional 
array of identical pixels or a two dimensional array of
identical pixels which range from 1 to $P$ (Figure 1).
For spectrally broad light, the fringe can be dispersed
in the cross fringe direction with the use of a two dimensional
detector.
For a focal plane interferometer of this type,
the influence of source shot noise on 
fringe phasor estimation has
been fully formulated (\cite{wal73}, \cite{goo85}). 
Let the interferometer be illuminated by a source and
spatially smooth background.
The intensity pattern on the  $r$th detector ($r$th baseline)
is given by

\begin{equation}
<I_r({\bf x})> = 2<I_0> \Bigl[1+\gamma_r
                 \cos(\kappa{\bf x}\cdot {\bf B}_r/d + \phi_r)\Bigr],
\label{fpint}
\end{equation}

where $<I_0>$ is the average intensity in each subbeam at the
detector, ${\bf B}_r$ is the baseline vector, 
$\kappa=2\pi/\lambda$ is the wave number, and $d$ is 
the distance between the aperture plane and the detector,
${\bf x}$ is the spatial vector in the detector plane,
and $\gamma_r \exp(i\phi_r)$ is the complex visibility function
at the baseline vector ${\bf B}_r$. In deriving (\ref{fpint}), we have
assumed that the incident light is spectrally narrow so that
the fringe visibility depends on only the spatial correlation of
the field. 

Because of the presence of background illumination,

\begin{equation}
I_0 = I_0^s + I_0^b,
\end{equation}

where $I_0^s$ and $I_0^b$ are source and background intensities
respectively. Let $\gamma^s_r$ be the fringe visibility of the source
in the absence of the background, then

\begin{equation}
\gamma_r = \frac{I_0^s}{I_0^s + I_0^b} \gamma^s_r.
\end{equation}

The complex visibility function $\gamma^s_r \exp(i\phi_r)$
is determined by the $uv$ coordinates of the $r$ th baseline $(u,v)$ 
and the source
brightness distribution on the sky $S(x,y)$ by a Fourier transform relation:

\begin{equation}
\gamma^s_r \exp(i\phi_r)
=\frac{\int S(x,y)\exp\{-2 \pi i(ux+vy)\} dx dy}
      {\int S(x,y) dx dy}.
\end{equation}

In an effort to reduce the clutter in the equations we henceforth
drop the vector notation, but bear in mind that spatial frequencies,
pixel locations, etc. are really vectors.
The photoelectron detection theory 
(\cite{wal73}) takes into account the discrete
nature of both photons and detector pixels.  
The average 
photoelectron count $<k_r(p)>$ at the pixel location specified by
the integer index $p$ of the detector is proportional to
$<I(x)>$: 
 
\begin{equation}
<k_r(p)> = 2<K_0>\Bigl[1+\gamma_r\cos(p\omega_r + \phi_r)\Bigr].
\label{kp}
\end{equation}

Here, $<...>$ denotes averaging over the photoelectron-detection
process. The product $p\omega_r$ is understood to be the scalar
product of the pixel position vector ${\bf p}$ and the spatial
frequency vector ${\bf \omega}_r$ expressed in inverse pixel units.

Let $<C>$ be the average number of photoelectrons detected by
the entire array in one integration period, and let 2$<N>$ be
the average number of photoelectrons per detector per integration
time. Clearly then, $<C> = 2<N>n_b$, and thus $<N> = 
<C>/\{n(n-1)\}$.  According to (\ref{kp}), the average number of
photoelectrons per detector is equal to $2<K_0>P$, and thus
$<K_0>P = <N>$.

\subsection{Inversion without total count}

An analysis based on the photodetection theory gives
the mean image $I_1$ as follows (NM01).

\begin{eqnarray}
I_1(q) & = & <N>\sum_r \gamma_r \cos(\omega_r q + \phi_r) \nonumber \\
& = & <N>\sum_r \frac{I^s_0}{I^s_0+I^b_0}\gamma_r^s \cos(\omega_r q + \phi_r),
\end{eqnarray}

where  $q$ refers to the pixels in the synthesized image; in
particular $q$  ranges from -Q/2 to +Q/2.

The variance $V[i_1(q)]$ in the synthesized image $i_1(q)$
is given by

\begin{eqnarray}
V[i_1(q)] 
& = & n_b (<N>+\frac{P\sigma^2}{2}) 
  = \frac{<C>}{2} + n_b \frac{P\sigma^2}{2}.
\end{eqnarray}

For a point source ($\gamma^s_r = 1$) at the phase center ($\phi_r
=0$), the SNR of the synthesized image is

\begin{equation}
\frac{I_1(0)}{\sqrt{V[i_1(0)]}}
=\frac{ (<C>/2) \{ I_0^s/(I_0^s+I_0^b)\} }
 {\sqrt{ <C>/2 + n_bP\sigma^2/2 }}.
\end{equation}

\subsection{Inversion with total count}

An analysis which takes into account shot noise and detector read
noise gives
the mean image $I_2$ as follows 

\begin{eqnarray}
I_2(q) & = & <N>\sum_r [\gamma_r \cos(\omega_r q + \phi_r) +1] \nonumber \\
& = & <N>\sum_r [\frac{I^s_0}{I^s_0+I^b_0}\gamma_r^s \cos(\omega_r q +
\phi_r) + 1].
\end{eqnarray}

The variance $V[i_2(q)]$ in the synthesized image $i_2(q)$
is given by

\begin{eqnarray}
V[i_2(q)] & = & n_b (\frac{3<N>}{2}+\frac{3P\sigma^2}{4}) 
    + <N>\sum_r \gamma_r \cos(\omega_r q + \phi_r) \nonumber \\ 
 & = &\frac{3<C>}{4} + n_b \frac{3P\sigma^2}{4}
    + <N>\sum_r \gamma_r \cos(\omega_r q + \phi_r) .
\end{eqnarray}

This variance is newly derived for this work.
For a point source at the phase center, the SNR of the image is

\begin{equation}
\frac{I_2(0)}{\sqrt{V[i_2(0)]}}
= \frac{<N>\sqrt{n_b}(2I^s_0 + I^b_0)/(I^s_0 + I^b_0)}
{\sqrt{(3<N>/2 + 3P\sigma^2/4)+<N>I^s_0/(I^s_0+I^b_0)}}
\end{equation}

\section{$^n C_n$ Interferometer}

In the $^nC_n$ interferometer, all the $n$ beams interfere
on a single detector and $n_b$ fringes are superposed.
Both the baseline configuration and the detector can be
either one dimensional or two dimensional. 
As an example, a two dimensional 
$^3C_3$ interferometer is schematically presented
in Figure 2. 
A special case
is the combination of a one dimensional baseline configuration
and a two dimensional detector for which the superposed fringes
are dispersed in the cross fringe direction so that
spectrally broad light can be used without bandwidth smearing (Figure 3). 
We call an $^nC_n$ interferometer of this type an $^nC_n^\prime$
interferometer.

\placefigure{fig2}

\placefigure{fig3}

Let the interferometer be composed of $n$ identical apertures
and let it be illuminated by a source and spatially smooth
background.
The classical intensity distribution of the interference
pattern by the $n$ apertures has the average value

\begin{equation}
<I({\bf x})> = <I_0> \Bigl[ n + 2 \sum_{g<h} \gamma_{gh}
\cos(\kappa{\bf x}\cdot {\bf B}_{gh}/d + \phi_{gh}) \Bigr],
\end{equation}

where the various symbols have meanings similar to those in (\ref{fpint}).
$gh$  denotes the baseline $gh$ corresponding to the apertures 
$g$ and $h$. Let $<k({\bf p})>$ denote the photoelectron count
distribution due to $<I({\bf x})>$. As in \S2, we discontinue
the vector notation, assume that the total number of pixels is $P$,
and note that $<k(p)>$ is proportional to $<I(x)>$:

\begin{equation}
<k(p)> = <Q_0> \Bigl[n + \sum_{g<h}\gamma_{gh}
\cos(p\omega_{gh} + \phi_{gh}) \Bigr].
\end{equation}

Here $<Q_0>$ has approximately the same meaning as $<K_0>$ in
\S2. However, since there is no beam splitting,
 $<Q_0> = (n-1)<K_0> $. 
We also define $<M> = P<Q_0> = <C>/n$.

\subsection{Inversion without total count}

We find
the mean synthesized image $I_3$ to be

\begin{eqnarray}
I_3(q)    &  = & <M> \sum_{i<j} \gamma_{ij}\cos(q\omega_{ij}+\phi_{ij}).
\end{eqnarray}

The variance $V[i_3(q)]$ in the synthesized image $i_3(q)$ is given by

\begin{eqnarray}
V[i_3(q)] 
& = &  \frac{n<M>+P\sigma^2}{2}n_b+<M>(n-2)
\sum_{i<j}\frac{I^s_0}{I^s_0+I^b_0}\gamma^s_{ij}
\cos(q\omega_{ij}+\phi_{ij})
.
\end{eqnarray}

This variance is derived in NM01.
For a point source at the phase center, the SNR of the image
is

\begin{eqnarray}
\frac{I_3(0)}{\sqrt{V[i_3(0)]}}  & = &
\frac
{
<M> \{I_0^s/(I_0^s+I_0^b)\} \sqrt{n_b}
}
{\sqrt{ (n<M>+P\sigma^2)/2+<M>(n-2)
\{I_0^s/(I_0^s+I_0^b)\}} 
} \nonumber \\
& = &
\frac{<C>\{I_0^s/(I_0^s+I_0^b)\}\sqrt{(n-1)/n} }
 {\sqrt{<C>+P\sigma^2 + <C>\{I_0^s/(I_0^s+I_0^b)\}
     \{2(n-2)/n\} }}. 
\end{eqnarray}

\subsection{Inversion with total count}

We find
the mean synthesized image $I_4$ to be

\begin{eqnarray}
I_4(q)    &  = & <M>[\frac{n}{2}+
\sum_{i<j} \gamma_{ij}\cos(q\omega_{ij}+\phi_{ij})] \\
  & = & \frac{<C>}{2}+
\frac{<C>}{n}\sum_{i<j} \gamma_{ij}\cos(q\omega_{ij}+\phi_{ij}).
\end{eqnarray}

The variance $V[i_4(q)]$ in the synthesized image $i_4(q)$ is given by

\begin{eqnarray}
V[i_4(q)] &  = & \frac{<M>}{2}\left[n(n_b+\frac{1}{2}) + 
2(n-1)\sum_{i<j} 2(j-1) \gamma_{ij}\cos(q\omega_{ij}+\phi_{ij})\right] \\
          &   & + P\sigma^2\left[\frac{n_b}{2}+\frac{1}{4}\right].
\end{eqnarray}

This variance is newly derived for this work.
For a point source at the phase center, the SNR of the image is

\begin{equation}
\frac{I_4(0)}{\sqrt{V_4(0)}} =
\frac{<C>n}{\sqrt{<C>(3n^2-5n+3)+P\sigma^2(n^2-n+1)}}.
\end{equation}

\section{Limiting Cases for Spectrally Broad Light}

In this section, we consider the source shot noise limit,
background shot noise limit, and detector read noise limit
for a point source and an extended source.
For each case, the results of inversion with and without
total count are shown.

Formulae for means, variances, and SNRs were obtained
for spectrally narrow light in the previous section. 
For a two-dimensional $^nC_n$ interferometer, the maximum spectral
bandwidth is limited to about $D/B$ where $D$ is the aperture
diameter and $B$ is the maximum baseline length.
If we use a bandwidth broader than this, fringes will be smeared out.
In order to disperse fringes for a general two-dimensional
baseline configuration, pupil remapping which transforms the
exit pupil to a $^nC_n^\prime$-type interferometer is necessary. 
The effect of pupil remapping on the signal-to-noise ratio analysis
is beyond the scope of this paper. However we can imagine that
the SNR of the one-dimensional $^nC_n^\prime$ interferometer
is a good measure of the SNR of an interferometer with pupil remapping. 
For $^nC_2$ and $^nC_n^\prime$ interferometers, fringes can be
dispersed in the cross fringe direction.
The formulae for SNRs obtained for
spectrally narrow light hold also for dispersed fringes as can
be easily verified as follows.

We here consider the case with the $^nC_2$ interferometer and
inversion without total count.  Let $P_x$ be the number of
pixels in the fringe direction and $P_y$ be that in the cross
fringe direction. Therefore $P = P_x P_y$. We assume that
the photon spectrum within the bandwidth is not steep and the number
of photons in one detector row is given by $<C>/P_y$.
Then the SNR of an image synthesized from a set of fringes in
detector rows corresponding to $1/P_y$ of the spectral bandwidth
is given by 

\begin{equation}
snr = \frac{<N>/P_y \sum_r \gamma_r \cos(\omega_r q + \phi_r)}
      {\sqrt{n_b(<N>/P_y + P_x\sigma^2/2)}}.  
\end{equation}

If we further assume that the total bandwidth is narrow enough
so that the source structure does not change within it. In this case,
$P_y$ images can be coadded to form a final image. The SNR of
the final image is given by

\begin{eqnarray}
SNR & = & \sqrt{P_y} snr \nonumber \\
    & = & \frac{<N> \sum_r \gamma_r \cos(\omega_r q + \phi_r)}
      {\sqrt{n_b(<N> + P_xP_y\sigma^2/2)}} \nonumber \\
    & = & \frac{<N> \sum_r \gamma_r \cos(\omega_r q + \phi_r)}
      {\sqrt{n_b(<N> + P\sigma^2/2)}}.
\end{eqnarray}

This happens because the photon count $<N>$ or $<C>$ and the 
pixel number $P$ are linear in the expressions of
the mean and variance. The situation is the same for the other 
SNR expressions.

\subsection{Numbers  of Pixels}

Let the central frequency be $\nu$ and the bandwidth be $\Delta\nu$.
Then there are 2$\nu/\Delta\nu$ fringe cycles in each interferogram.
For the Nyquist sampling theorem, 4$\nu/\Delta\nu$ pixels are
needed in the fringe direction. 

For the $^nC_2$ and $^nC_n^\prime$ 
interferometers, we consider the number of pixels in the cross
fringe direction (dispersion direction) in each dispersed fringe.
The fractional bandwidth $\Delta\nu^*/\nu$ for which bandwidth
smearing is small is given by $\Delta\nu^*/\nu < D/B$.
If we require that $\Delta\nu^*/\nu = D/(2B)$, bandwidth smearing
will be negligible. For a spectral bandwidth of $\Delta\nu$,
$\Delta\nu/\Delta^*\nu = 2(\Delta\nu)/(D/B)$ pixels are needed
as the number of pixels in the cross fringe direction.
So the number of pixels of the two dimensional detector
is $P = 4\nu/\Delta\nu \times 2(\Delta\nu)/(D/B) = 8(B/D)$.

For the $^nC_n$ interferometer, the longest baseline can be
aligned to one of the two sides of a two dimensional detector
and the minimum number of pixels is
$ P = (4\nu/\Delta\nu)^2 = 16(B/D)^2$ and the maximum allowed
fractional bandwidth is $D/B$.

\subsection{SNR Tables}

SNR expressions of $^nC_2$, $^nC_n$, and $^nC_n^\prime$
interferometers are given for a point source and an extended source,
for source shot noise limit, background shot noise limit, and
detector read noise limit, and for inversion with and without
total count in Tables 1 through 5.
In some cases the SNR expressions are not well defined and omitted
from the tables.

We define a quantity
$F_\nu$ which is proportional to the photon spectrum of the 
source, collecting area, and integration time.
We express the total photon count for the  
$^nC_2$ and $^nC_n^\prime$ interferometers as 
$<C> = F_\nu \Delta\nu$. Then the total photon count for the two-dimensional
$^nC_n$ interferometer is $<C> = F_\nu \nu (D/B)$.
The SNRs are expressed in terms of $F_\nu$, $\Delta\nu$,
and $D/B$.

\subsection{Point Source/Source Shot Noise Limit}

The case with a point source in the shot noise limit is given
in Table 1.

By taking ratios of the SNR expressions, we find

\begin{equation}
SNR_1/SNR_3 =
\sqrt{\frac{3n-4}{2n-2}}\sqrt{\frac{\Delta\nu/\nu}{D/B}},
\end{equation}

\begin{equation}
SNR_1/SNR_3^\prime =
\sqrt{\frac{3n-4}{2n-2}} \geq 1,
\end{equation}

\begin{equation}
SNR_2/SNR_4 =
\sqrt{\frac{4(3n^2-5n+3)}{5n^2}}\sqrt{\frac{\Delta\nu/\nu}{D/B}},
\end{equation}

\begin{equation}
SNR_2/SNR_4^\prime =
\sqrt{\frac{4(3n^2-5n+3)}{5n^2}} \geq 1,
\label{snr24}
\end{equation}

\begin{equation}
SNR_2/SNR_1 =
\sqrt{\frac{8}{5}}.
\label{snr21}
\end{equation}

In a typical situation, $\Delta\nu/\nu > D/B$ and 
the $^nC_2$ and $^nC_n^\prime$ interferometers are superior to
the two-dimensional $^nC_n$ interferometer. 
As the expression (\ref{snr24})
shows,  the $^nC_2$ interferometer is superior to 
the  $^nC_2^\prime$ interferometer by a factor $\sqrt{12/5}$ in
SNR for a large $n$ for inversion with total count. 
In the source shot noise limit, the total count information is
genuinely of the source. The highest SNR is obtained for
the $^nC_2$ interferometer for inversion with total count 
(eq. (\ref{snr21})).

\subsection{Background Limit}

In the background limit, only inversion without total count
is the valid method of analysis, since the total count is
dominated by background photon counts. The SNR expressions are given
in Table 2.

By taking the ratios, we find

\begin{equation}
SNR_1/SNR_3 = \sqrt{\frac{n}{2(n-1)}}\sqrt{\frac{\Delta\nu/\nu}{D/B}},
\end{equation}

and

\begin{equation}
SNR_1/SNR_3^\prime = \sqrt{\frac{n}{2(n-1)}} \leq 1.
\end{equation}

As in the case with the source shot noise limit,
the two-dimensional $^nC_n$ interferometer has a disadvantage of 
the narrow bandwidth. For a large $n$, the $^nC_n^\prime$
interferometer
is slightly superior to the $^nC_2$ interferometer by a factor
$\sqrt{2}$ in SNR.

\subsection{Read noise limit}

In the read noise limit, the numbers of pixels 
for the $^nC_2$ and $^nC_n^\prime$ interferometers are
$P = 8(B/D)$ and that for the two-dimensional $^nC_n$ interferometer
is $P = 16 (B/D)^2$. The SNR expressions are given in Table 3.

If we take the ratios of the SNRs, we find

\begin{equation}
SNR_1 / SNR_3 = \frac{\sqrt{2}}{n-1}\frac{\Delta\nu}{\nu}
(\frac{B}{D})^{3/2},
\end{equation}

\begin{equation}
SNR_1 / SNR_3^\prime = \frac{1}{n-1} \leq 1,
\end{equation}

\begin{equation}
SNR_2 / SNR_4 = \frac{4}{n}\sqrt{\frac{n^2-n+1}{3(n^2-n)}}
\frac{\Delta\nu}{\nu}(\frac{B}{D})^{3/2},
\end{equation}

\begin{equation}
SNR_2 / SNR_4^\prime = \frac{2}{n}\sqrt{\frac{2(n^2-n+1)}{3(n^2-n)}}
\leq 1.
\end{equation}

The $^nC_n^\prime$ interferometer is always superior to 
the $^nC_2$ interferometer because of the smaller number of
detector pixels. 
The two-dimensional $^nC_n$ interferometer has a limited
use for a small $B/D$.

\subsection{Extended Source/Source Shot Noise Limit}

For a fully extended source, there is no signal other than
the total count.
Only inversion with total count is the valid method of analysis
for the fully extended source. The SNR expressions are given
in Table 4.

By taking the ratios, we find

\begin{equation}
SNR_2 / SNR_4 = \sqrt{\frac{n^2-n+1}{3}}
\sqrt{\frac{\Delta\nu}{\nu}\frac{B}{D}},
\end{equation}

\begin{equation}
SNR_2 / SNR_4^\prime = \sqrt{\frac{n^2-n+1}{3}} \geq 1.
\end{equation}

The $^nC_2$ interferometer is superior to 
the $^nC_n$ and $^nC_n^\prime$ interferometers.
%
%

\subsection{Extended Source/Background Limit}

For an extended source, there is no signal other than
the source total count. However, the total count is
dominated by background photons. Therefore 
neither inversion without total count nor inversion
with total count is the valid method of analysis.

\subsection{Extended Source/Read Noise Limit}

As in the case with extended source/shot noise limit,
the only valid method of analysis is inversion with total count.
The SNR expressions are given in Table 5.

By taking ratios, we find

\begin{equation}
SNR_2/SNR_4 = \sqrt{\frac{4(n^2-n+1)}{3(n^2-n)}}\frac{\Delta\nu}{\nu}
(\frac{B}{D})^{3/2},
\end{equation}

and 

\begin{equation}
SNR_2 / SNR_4^\prime = \sqrt{\frac{2(n^2-n+1)}{3(n^2-n)}} \leq 1.
\end{equation}

The two-dimensional $^nC_n$ interferometer is useful only for
a small $B/D$. For large $n$, $SNR_2/SNR_4^\prime$ attains
an asymptotic value of $\sqrt{2/3}$.

\section{Sensitivities of Interferometers Composed of Nine 10-m
telescopes}

In this section, we calculate the sensitivities of $^nC_2$ and
$^nC_n^\prime$ interferometers composed of nine 10-m
telescopes and compare them with the sensitivity of
a 30-m single telescope with adaptive optics.
10-$\sigma$ detection limits for a point source
are obtained for one-hour integration and plotted in Figure 4.

\placefigure{fig4}

\subsection{J and H}

Throughput of the interferometers is assumed to be 10\%,
while that of the single telescope with adaptive optics
is assumed to be 60\%. The read noise level is assumed to
be 10 electrons per pixel and the maximum baseline length
which determines the number of detector pixels is assumed to be 200 m.
The coherent integration time is assumed to be one second, while
a larger value will decrease the effect of detector read noise.

It turned out that both read noise
and sky background due to OH airglow affect the performance of
the interferometers. We evaluate the SNRs for inversion without
total count because sky background is much greater than the signal
near the detection limit.
The sensitivity of the telescope is limited
by sky background.  At J, limiting magnitudes are 24.7, 25.7, and 28.0
respectivity for the $^nC_2$ interferometer, $^nC_n^\prime$ interferometer,
and the single telescope. At H, they are 23.6, 24.0, and 26.2 respectively.

\subsection{Thermal Infrared Wavelengths}

For the interferometers, we evaluate the SNRs for
inversion without total count because the instrument background
is much stronger than the source signal near the detection limits.
The temperature of interferometer optics is assumed to be 273 K
and the optical throughput is assumed to be 10\%.
Therefore the background emission is estimated from
a 273K blackbody with 90\% emissivity.
For the single telescope with adaptive optics, the throughput of
the optics is assumed to be 60\%.  
The temperature of the 
optics is assumed to be 273 K,  and its emissivity is assumed
to be 40\%.  

Longward of 2 $\mu$m, the sensitivities
do not depend on beam combination geometry very much.
This implies that detector read noise is negligible.
At 2.2 $\mu$m, the sensitivities of the interferometers are
around  330 nJy or 22.9 mag, while that of the single
telescope is 10 nJy or 26.9 mag. Longward of 5 $\mu$m,
the single telescope is 5.8 mag more sensitive than
the interferometers. At 10.5 $\mu$m, the sensitivities
of the interferometers are about 2 mJy or 10.7 mag,
while that of the single telescope is 57 $\mu$Jy or 14.4 mag.
At wavelengths shorter than 2 $\mu$m, the baseline length
affects the sensitivities of the interferometers due to
the change in the number of pixels, while longward of
2 $\mu$m, it does not matter as long as
the source is unresolved.
Here we do not consider the case for a fully extended objects.
As we have seen from the previous section, the interferometers
are not sensitive to fully extended objects under background
limited conditions.

\subsection{Loss-Free Interferometers}

In the above evaluation of sensitivities, it is assumed that
the throughput of the interferometers is 10\%, while that
of the single telescope is 60\%. It is natural to question
whether the difference in sensitivities is due to that in
throughput or something more fundamental.

As discussed by PK89, there is a basic difference between an interferometer
and a single telescope in that the former takes the correlation of signal
from multiple apertures. However the loss of SNR
by this process is only a factor of $\sqrt{2}$ in the case of an ideal
$^nC_2$ interferometer with inversion without total count in
the source shot noise limit. A similar situation is expected in
the case of background shot noise limit where the number of detector
pixels does not affect the SNR. Then the only situation in which
the difference is conspicuous is the read noise limit where
the interferometer requires many more detector pixels to sample fringes.
It is of theoretical interest to assume ideal interferometers
and compare their performance with that of the single telescope.

Again we compare the interferometers with nine 10-m telescopes
and a 30-m single telescope. Since each of the 10-m telescope
and the 30-m single telescope require adaptive optics,
the throughputs are assumed to be 60\%.
The sensitivity limits are plotted in Figure 5, where the difference
is about by a factor of five.
Therefore the fundamental difference between in the interferometers
and the single telescope 
is rather small and the large
difference seen in the previous analysis is due to that in throughput.

\placefigure{fig5}

\subsection{Baseline Dependence}

The SNR expressions of the interferometers are not
explicitly baseline dependent.
Since the astronomical source is assumed to be unresolved,
only the total number of detector pixels is dependent on 
the lengths of baselines.  Detector read noise is more
significant as the baselines become longer. 
In order to use long baselines, it is essential to keep detector read
noise low so that the SNR is not limited by read noise.

For a source of finite extent,
long baselines start to resolve it and the visibilities $\gamma^s_r$
of those baselines decrease from unity and the phases $\phi_r$
become less coherent depending on the symmetry of the source.
For the  extended source, the SNRs of the interferometers
are calculated as follows. For a particular baseline configuration,
the complex visibilities are obtained by the Fourier transform
relation (4) and the number of detector pixels are calculated 
as above. By inserting those numbers and taking the summation over
different baselines,
one can obtain the SNRs of the dirty images.

\section{Discussion}

\subsection{Interferometry for extragalactic astronomy}
 
One of the motivations of this analysis is to evaluate 
the capability of interferometry for extragalactic astronomy.
We inevitably have to consider extended objects for this purpose
and take into account the guide star availability.
Fringe tracking requires the presence of a natural guide star 
in the infrared. 
If the adaptive optics also needs the same natural guide star,
the limiting magitude is around R = 15, and an isoplanatic angle
is order of 10$^{\prime\prime}$ at 2 $\mu$m. 
Requirements for a natural guide star limit the range of
observing targets to bright active galactic nuclei (AGNs) and 
bright quasars.  

The central continuum source of an AGN or a quasar will act as
a natural guide star and the fringe visibilities are close to
unity and phases are close to zero. Therefore SNR calculations
made in the previous section are applicable to this case.
Small deviations in 
a reconstructed image from a point source indicate the
presence of a faint extended
structure such as jet. The 10$\sigma$ limiting magnitude of K=25 indicates
SNR = 10$^5$ for a K=15 quasar. In other words, jet can be searched
for with a dynamic range of 10$^5$:1.

Now we consider resolving broad-line regions (BLRs). 
According to the photoionization model (\cite{dav79}),
the size of a BLR scales as $\sqrt{L}$ where $L$ is the luminosity
of the AGN. 
For an AGN or a nearby quasar toward which the
geometry is Euclidean, the observed flux $f$ is given by
$f = L/d^2$ where $d$ is the distance to the object.
Then the angular size $\theta \propto \sqrt{f}$. Therefore
the brightest quasar has the largest BLR in angular size.
3C273 is the brightest quasar (K=10) and easiest to observe because
Pa$\alpha$ is redshifted to K band. The necessary spatial resolution
is order of 1 milliarcsecond or the necessary
baseline length is 400 m. The FWHM of the line is 3400 kms$^{-1}$
and the line/continuum ratio below 8000 kms$^{-1}$ is 0.08
(\cite{sel83}). In order to obtain the spatial structure of the BLR,
we wish to spectrally resolve the broad line or at least separate the
blue and red components. If we require a spectral resolution of 3000
kms$^{-1}$, the continuum flux per spectral resolution is K=14
equivalent. A spectral image will be composed of an unresolved source
of K=14 and a resolved source with 8\% of the flux of the unresolved source.
Again, the complex visibilities are close to those of a point source
and the SNR argument made above for jet applies to this case as well.
We expect that the BLR of 3C273 will be resolved with a dynamic range more than
10$^5$:1.
At least for AGNs and quasars, infrared interferometry with large
telescopes appears effective.

\subsection{Comments on RR99}

In this subsection, we compare our results with that of RR99 on
Fizeau-type interferometers or two-dimensional $^nC_n$
interferometers.
First of all, for spectrally broad light, we have seen
that two-dimensional $^nC_n$ interferometers or
Fizeau-type interferometers have limitations in
the ratio of the telescope aperture diameter and
the baseline length to avoid bandwidth smearing.
This is not discussed in RR99. 
Second, RR99 states 
that the $^nC_n$ interferometer is superior to 
the $^nC_2$ interferometer. 
However, we have seen from the analysis of limiting
cases that the situation is not so simple.
As PK89 noted, there is no major difference in performance
between both interferometers in the source shot noise limit.
This is also true with the case for the background limit.
In the read noise limit, the $^nC_n^\prime$ interferometer
is superior to the $^nC_2$ interferometer due to the difference
in the number of detector pixels.    

These differences originates from that in the definitions
of the SNRs. RR99 defines the SNR as that of the central
peak in the interferogram directly formed by beam combination.
On the other hand, we define the SNR as that of the dirty
image formed by aperture synthesis. In the latter,
the interferogram formed by beam combination is Fourier transformed
to measure complex visibilities which fill the $uv$ plane
and then they are  Fourier transformed back to the image plane to form
a dirty image. In RR99, the detector pixels of concern at the
focal plane are only a few pixels at which the peak is located,
while in the latter many pixels which sample the entire
interferogram are considered.

\subsection{Remaining Issues}

We have obtained the sensitivity limits of the interferometers
for perfect wavefront correction by adaptive optics and
perfect fringe tracking. In reality, both adaptive optics and
fringe tracking require a guide star and there is a limitation
in the precision of wavefront and delay compensations.
Imperfection in phase directly influences the fringe visibility
$\gamma$ which is proportional to the SNRs. 
An analysis which takes into account this imperfection in phase
is difficult because introduction of atmospheric disturbances
opens up a vast parameter space.  Our analysis corresponds only
to the best case designing.

In this paper, we considered only nonredundant baselines.
However, if we consider a closely packed array of telescopes,
we have to take into account the redundancy of the baselines.
Although the behavior of SNRs are expected to be the same
as the nonredundant case, deriving formulae of SNRs for
the redundant case is non trivial or analytically impossible.

Finally one essential and difficult question is
whether the definition of the SNR as that of the dirty image
is valid. In other words, we have left out the
issue of deconvolution by assuming that it is perfect.
We believe that this is practically of no problem because
the SNR of the dirty beam is usually much higher than that
of the dirty image.
Nevertheless deconvolution is a nonlinear process and the error propagation
through it is not fully understood. 

\section{Conclusions}

We present the formulae of SNRs of interferometers for the
purpose of comparing ground-based interferometers composed of
multiple large telescopes with a single giant telescope with
adaptive optics.
In deriving sensitivity limits, perfect wavefront correction 
by adaptive optics and perfect cophasing
by fringe tracking are assumed.
We consider the case in which $n$ beams are pairwise combined
at $n(n-1)/2$ detectors ($^nC_2$ interferometer)
and the case in which all the $n$ beams
are combined at a single detector ($^nC_n$ interferometer).

\noindent (1)
In the read noise limit, the $^nC_n$ interferometer with 
one dimensional baseline configuration is superior to the $^nC_2$
interferometer, while in the background limit, the advantage of
the $^nC_n$ interferometer is small.

\noindent (2)
We compare the 
point-source sensitivities of interferometers composed of nine
10-m diameter  telescopes and a 30-m diameter single telescope
with adaptive optics between 1 and 10 $\mu$m
for 10-$\sigma$ detection in one hour.  
Between 1 and 2 $\mu$m, the sensitivities of the interferometers are
limited by detector read noise and sky background, while longward of 2 $\mu$m,
they are limited by the instrumental thermal background.
At K band, the 
sensitivities of the interferometers are around 
23 mag, while that of 
the single telescope is 27 mag.



\acknowledgments

I thank H. Matsuhara for allowing me to use figures
and the referee P. Lawson for useful comments.

\clearpage
 
\begin{deluxetable}{ccc}
\footnotesize
\tablecaption{SNR (Point Source / Shot Noise Limit) \label{tbl-1}}
\tablewidth{0pt} 
\tablehead{
\colhead{Case} & \colhead{Notation} & \colhead{Expression}
}
\startdata
$^nC_2$/inversion w/out total count 
& $SNR_1$ & $\sqrt{\frac{F_\nu \Delta\nu}{2}}$  \nl
$^nC_2$/inversion with total count 
& $SNR_2$ & $\sqrt{\frac{4}{5}}\sqrt{F_\nu \Delta\nu}$ \nl
$^nC_n$/inversion w/out total count 
& $SNR_3$ &  $\sqrt{\frac{n-1}{3n-4}}\sqrt{F_\nu \nu\frac{D}{B}}$ \nl
$^nC_n$/inversion with total count 
& $SNR_4$ &  $\sqrt{\frac{n^2}{3n^2-5n+3}}\sqrt{F_\nu \nu\frac{D}{B}}$ \nl
$^nC_n^\prime$/inversion w/out total count
& $SNR_3^\prime$ 
& $\sqrt{\frac{n-1}{3n-4}}\sqrt{F_\nu \Delta\nu}$ \nl
$^nC_n^\prime$/inversion with total count
& $SNR_4^\prime$
&  $\sqrt{\frac{n^2}{3n^2-5n+3}}\sqrt{F_\nu \Delta\nu}$ \nl
\enddata
\end{deluxetable}

\begin{deluxetable}{ccc}
\footnotesize
\tablecaption{SNR (Point Source / Background  Limit) \label{tbl-2}}
\tablewidth{0pt} 
\tablehead{
\colhead{Case} & \colhead{Notation} & \colhead{Expression}
}
\startdata
$^nC_2$/inversion w/out total count & $SNR_1$ &
$\frac{I_0^s}{I_0^b} \sqrt{\frac{F_\nu\Delta\nu}{2}}$ \nl
$^nC_n$/inversion w/out total count & $SNR_3$ &
$ \frac{I_0^s}{I_0^b} \sqrt{\frac{n-1}{n}}\sqrt{F_\nu\nu\frac{D}{B}}$ \nl
$^nC_n^\prime$/inversion w/out total count & $SNR_3^\prime$ &
$\frac{I_0^s}{I_0^b} \sqrt{\frac{n-1}{n}}\sqrt{F_\nu\Delta\nu}$ \nl
\enddata
\end{deluxetable}

\begin{deluxetable}{ccc}
\footnotesize
\tablecaption{SNR (Point Source / Read Noise Limit) \label{tbl-3}}
\tablewidth{0pt} 
\tablehead{
\colhead{Case} & \colhead{Notation} & \colhead{Expression}
}
\startdata
$^nC_2$/inversion w/out total count & $SNR_1$ &
$\frac{1}{4\sqrt{n_b(B/D)}}\frac{F_\nu\Delta\nu}{\sigma}$ \nl
$^nC_2$/inversion with total count & $SNR_2$ &
$\frac{1}{\sqrt{6n_b(B/D)}}\frac{F_\nu\Delta\nu}{\sigma}$ \nl
$^nC_n$/inversion w/out total count & $SNR_3$ &
$\frac{1}{4}\sqrt{\frac{n-1}{n}} (\frac{D}{B})^2
\frac{F_\nu\nu}{\sigma}$ \nl
$^nC_n$/inversion with total count & $SNR_4$ &
$\frac{1}{4}\sqrt{\frac{n^2}{n^2-n+1}} (\frac{D}{B})^2
\frac{F_\nu\nu}{\sigma}$ \nl
$^nC_n^\prime$/inversion w/out total count & $SNR_3^\prime$ &
$ \frac{1}{\sqrt{8(B/D)}}\sqrt{\frac{n-1}{n}} 
\frac{F_\nu\Delta\nu}{\sigma}$ \nl
 $^nC_n^\prime$/inversion with total count & $SNR_4^\prime$ &
$\frac{1}{\sqrt{8(B/D)}}\sqrt{\frac{n^2}{n^2-n+1}} 
\frac{F_\nu\Delta\nu}{\sigma}$ \nl
\enddata
\end{deluxetable}

\begin{deluxetable}{ccc}
\footnotesize
\tablecaption{SNR (Extended Source/Source Shot Noise Limit) \label{tbl-4}}
\tablewidth{0pt} 
\tablehead{
\colhead{Case} & \colhead{Notation} & \colhead{Expression}
}
\startdata
$^nC_2$/inversion with total count & $SNR_2$ & 
$\sqrt{\frac{F_\nu\Delta\nu}{3}}$ \nl
$^nC_n$/inversion with total count & $SNR_4$ &
$\sqrt{\frac{F_\nu\nu(D/B)}{n^2-n+1}}$ \nl
$^nC_n^\prime$/inversion with total count & $SNR_4^\prime$ &
$\sqrt{\frac{F_\nu\Delta\nu}{n^2-n+1}}$ \nl
\enddata
\end{deluxetable}

\begin{deluxetable}{ccc}
\footnotesize
\tablecaption{SNR (Extended Source/Read Noise Limit) \label{tbl-5}}
\tablewidth{0pt} 
\tablehead{
\colhead{Case} & \colhead{SNR} & \colhead{Expression}
}
\startdata
$^nC_2$/inversion with total count & $SNR_2$ &
$\frac{1}{\sqrt{12(n^2-n)(B/D)}}\frac{F_\nu\Delta\nu}{\sigma}$ \nl
$^nC_n$/inversion with total count & $SNR_4$ & 
$\frac{1}{\sqrt{16(n^2-n+1)(B/D)^2}}\frac{F_\nu\nu(B/D)}{\sigma}$ \nl
$^nC_n^\prime$/inversion with total count & $SNR_4^\prime$ &
$ \frac{1}{\sqrt{8(n^2-n+1)(B/D)}}\frac{F_\nu\Delta\nu}{\sigma} $ \nl
\enddata
\end{deluxetable}

%
%






\figcaption[fig1.eps]{Schematic diagram of a focal plane interferometer.
There is one detector for each baseline. \label{fig1}}

\figcaption[fig2.eps]{Schematic diagram of a $^3C_3$ interferometer. Beams
from all apertures are superposed on a single detector. \label{fig2}}

\figcaption[fig3.eps]{Schematic diagram of a one-dimensional  $^3C_3$
interferometer. Fringes are spectrally dispersed in cross fringe direction.
 \label{fig3}}

\figcaption[fig4.eps]{Detection limits of the $^nC_2$- and 
1D $^nC_n$  
interferometers and that of the single giant telescope are
plotted as functions of wavelength. Symbols are sensitivities
are at J and H of $^nC_2$ (asterisk), 1D $^nC_n$ (square)
and telescope (circle). Assumed parameters are described in the text.
  \label{fig4}}

\figcaption[fig5.eps]{Loss-Free Interferometers. Throughputs of
the interferometers and the single telescope are assumed to be the
same. Meanings of the symbols are the same as in Figure 4.
\label{fig5}}


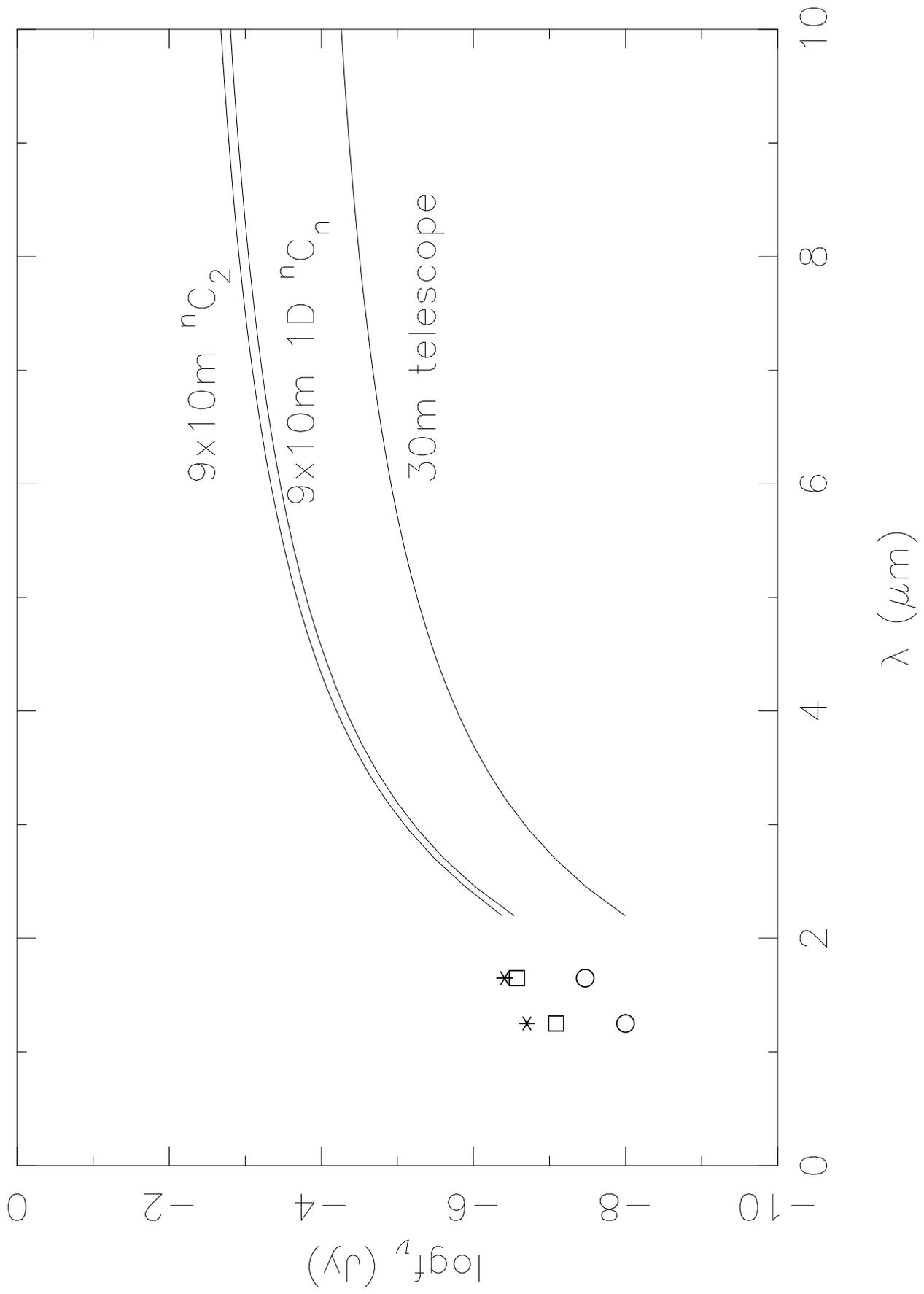
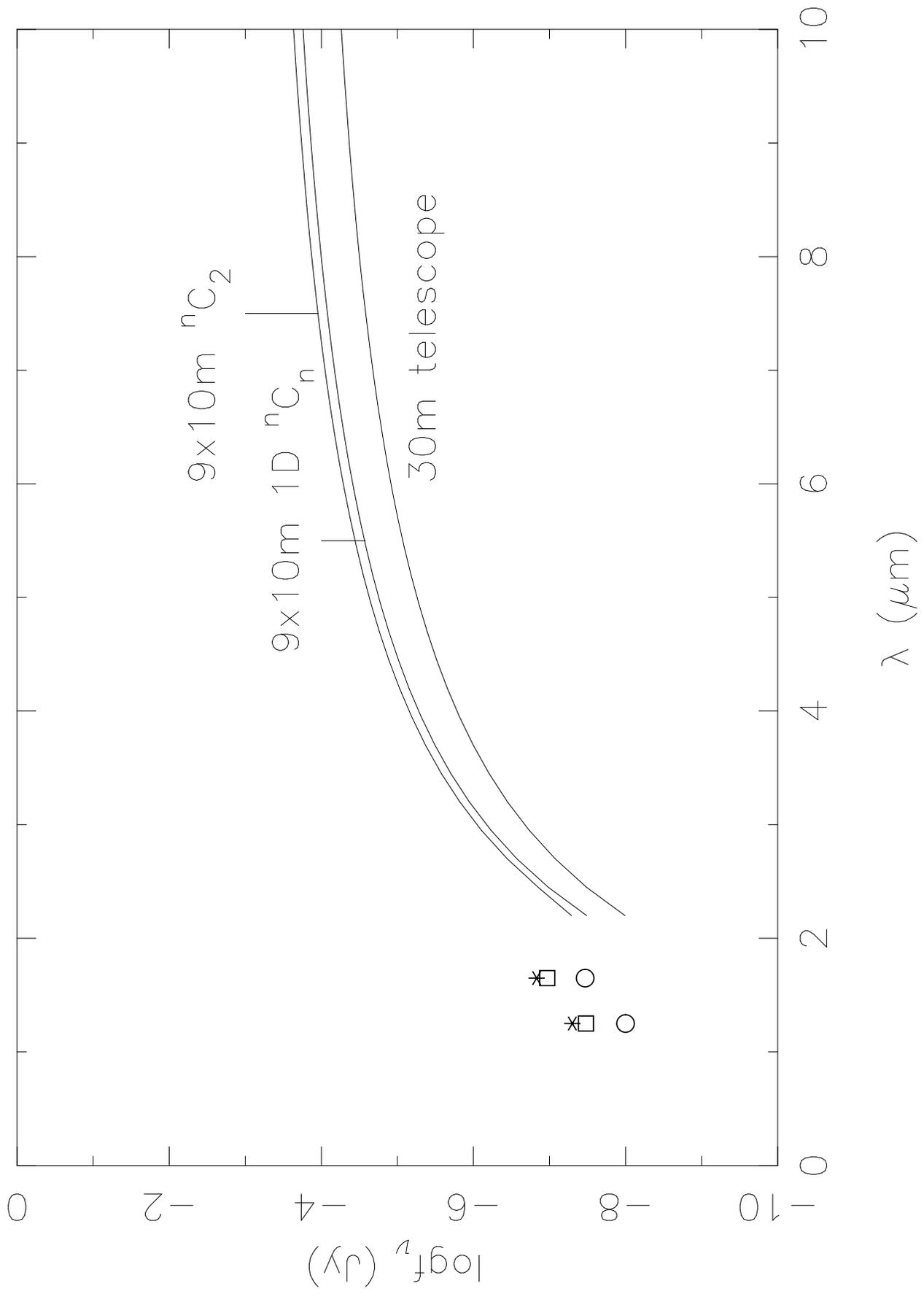
\end{document}